\documentclass[twocolumn,aps,prl,raggedbottom,nobalancelastpage,amssymb,superscriptaddress]{revtex4-1}

\usepackage{amsmath}
\usepackage{amsfonts}
\usepackage{graphicx}
\usepackage{appendix}
\usepackage{dsfont}
\usepackage{amssymb}
\usepackage{bm}
\usepackage{color}

\newcommand{\ave}[1]{\langle #1\rangle}
\newcommand{\ket}[1]{\left| #1\right\rangle}
\newcommand{\bra}[1]{\left\langle #1\right|}

\newcommand{\tb}{H_T^\mathcal{B}}

\newcommand{\td}{H_T^\mathcal{D}}
\newcommand{\tad}{H_T^\mathcal{A \dag}}
\newcommand{\tbd}{H_T^\mathcal{B \dag}}
\newcommand{\tcd}{H_T^\mathcal{C \dag}}
\newcommand{\tdd}{H_T^\mathcal{D \dag}}

\begin{document}

\title{Hanbury-Brown and Twiss interference of anyons}
\author{Gabriele Campagnano}
\affiliation{Department of Condensed Matter Physics, The Weizmann Institute of Science, Rehovot 76100, Israel}
\author{Oded Zilberberg}
\affiliation{Department of Condensed Matter Physics, The Weizmann Institute of Science, Rehovot 76100, Israel}
\author{Igor V. Gornyi}
\affiliation{Institut f\"{u}r Nanotechnologie, Karlsruhe Institute of Technology, 76021 Karlsruhe, Germany}
\affiliation{A.~F. Ioffe Physico-Technical Institute, 194021 St. Petersburg, Russia}
\author{Dmitri E. Feldman}
\affiliation{Department of Physics, Brown University, Providence, Rhode Island 02912, USA}
\author{Andrew C. Potter}
\affiliation{Department of Physics, Massachusetts Institute of Technology, Cambridge, Massachusetts 02139, USA}
\author{Yuval Gefen}
\affiliation{Department of Condensed Matter Physics, The Weizmann Institute of Science,  Rehovot 76100, Israel}

\date{\today}

\begin{abstract}
We present a study of an Hanbury Brown and Twiss (HBT)  interferometer realized with anyons. Such a device can directly probe entanglement and fractional statistics of initially uncorrelated particles. We calculate HBT cross-correlations of Abelian Laughlin anyons. The correlations we calculate exhibit partial bunching similar to bosons, indicating a substantial   statistical transmutation from the underlying electronic degrees of freedom. We also find qualitative differences between the anyonic signal and the corresponding bosonic or fermionic signals, indicating that anyons cannot be simply thought as intermediate between bosons and fermions.
\end{abstract}
\maketitle

Two-particle interference is a major pillar of quantum mechanics, very much like the phenomenon of single particle interference.  Such interference has been observed with photons in the historical Hanbury Brown and Twiss (HBT) experiment~\cite{HBT:1954,HBT:1956}, and much later  with electrons~\cite{Neder:2007}.  Quantum Hall systems can exhibit emergent particles (dubbed anyons) with fractional statistics~\cite{Leinaas:1977,Wilczek:1982}. Despite intensive study,   direct  signatures of anyonic  statistics remain elusive. Here we study an HBT interferometer with anyons, which can directly probe entanglement and fractional statistics of initially uncorrelated particles. Specifically, we calculate HBT cross-correlations of Abelian Laughlin anyons. The correlations exhibit partial bunching similar to bosons, indicating a substantial   statistical transmutation from the underlying electronic degrees of freedom~\cite{note1}. Furthermore, we find qualitative differences between the anyonic signal and the corresponding bosonic or fermionic signals, indicating that anyons cannot be simply thought as intermediate between bosons and fermions.

Edge channels of a fractional quantum Hall system offer a natural framework to study transport properties of anyons. Earlier attempts to  consider entanglement of such quasiparticles (QPs) either addressed  time-resolved correlation functions~\cite{Vishveshwara:2003} (which may be very hard to measure) or relied  on a single source geometry setup~\cite{Safi:2001,PhysRevLett.95.176402,PhysRevB.74.155324} (which may introduce superfluous interaction-induced correlations). Here we study zero frequency current-current correlations in a truly HBT interferometer setup, whose physics is governed by QPs dynamics. Because of their fractional charge and fractional statistics, scattering of these QPs  results in non trivial correlations. Below, we consider  the case $\nu=1/3$ for concreteness, but, our analysis can be generalized to other Laughlin fractions.

\begin{figure}[!ht] \begin{center}
\includegraphics[width=8cm]{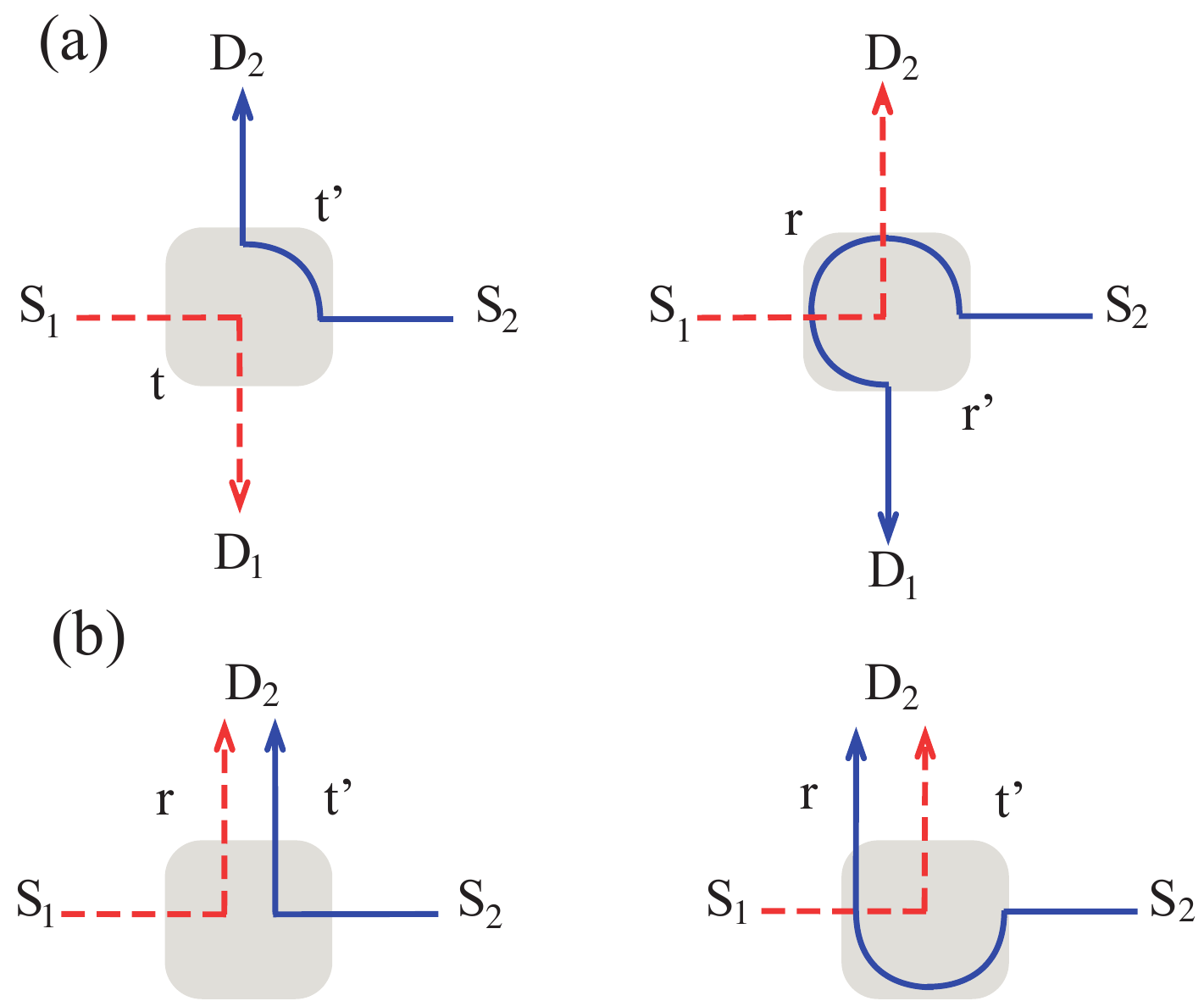}
\end{center}
\caption{Two-particle amplitudes contributing to: (a) P(1,1), the two particles are emitted from $S_1$ and $S_2$, and collected at $D_1$ and $D_2$; (b) P(0,2), both particles
are collected at $D_1$. $t,r,t^\prime,r^\prime$ are the single particle scattering amplitudes, $|t|^2+|r|^2=1,|t^\prime|^2+|r^\prime|^2=1$. Note the statistical factors, reflected by the winding of one particle around the other. For (a) they are $\text{exp}[i(1/2)\pi\nu]$ and $\text{exp}[i(3/2)\pi\nu]$, respectively.
 } \label{windings}
\end{figure}

Consider first a heuristic estimate of these correlations, outlined in Fig.~\ref{windings}. Two particles are emitted respectively from two sources $S_1$ and $S_2$ and scattered towards two detectors $D_1$ and $D_2$ by a beam splitter, e.g. a quantum point contact (QPC) for electrons and QPs, or a half silvered mirror for photons. We evaluate the probability $P(m,2-m)$, $m=0,1,2$, that $m$ particles are collected at the drain $D_1$ while ($2-m$) are collected at the drain $D_2$. Consider, e.g., the diagrams contributing to $P(1,1)$ [cf. Fig.~\ref{windings}(a)]. Each diagram represents an amplitude contributing to $P(1,1)$. Their weights are $t t^\prime \text{exp}[i(1/2)\pi\nu]$ and $r r^\prime \text{exp}[i(3/2)\pi\nu]$, with $\nu=0,1,1/3$ for bosons, fermions, and $\nu=1/3$ anyons, respectively. Note that we have included quantum statistics factors which reflect the extent by which one particle winds around the other. It follows that $P(1,1)=\left|t t^\prime \text{exp}[i(1/2)\pi\nu]+r r^\prime \text{exp}[i(3/2)\pi\nu]\right|^2$. For simplicity we consider symmetric scatterers, $|r|^2=|t|^2=1/2$, in which case $P(1,1)=(1/2)(1-\cos\pi\nu)$. Similarly, $P(2,0)=P(0,2)=(1/4)(1+\cos\pi\nu)$. For classical particles one sums up probabilities, rather than amplitudes, leading to $P_C(1,1)=1/2$, $P_C(2,0)=P_C(0,2)=1/4$. The results for fermions and bosons coincide with calculations based on second quantization~\cite{Blanter:2000}. For $\nu=1/3$ this results in boson-like bunching~\cite{Vishveshwara:2003} ($P_{\nu=1/3}(2,0)>P_C(2,0)$).

\begin{figure}[!ht] \begin{center}
\includegraphics[width=9cm]{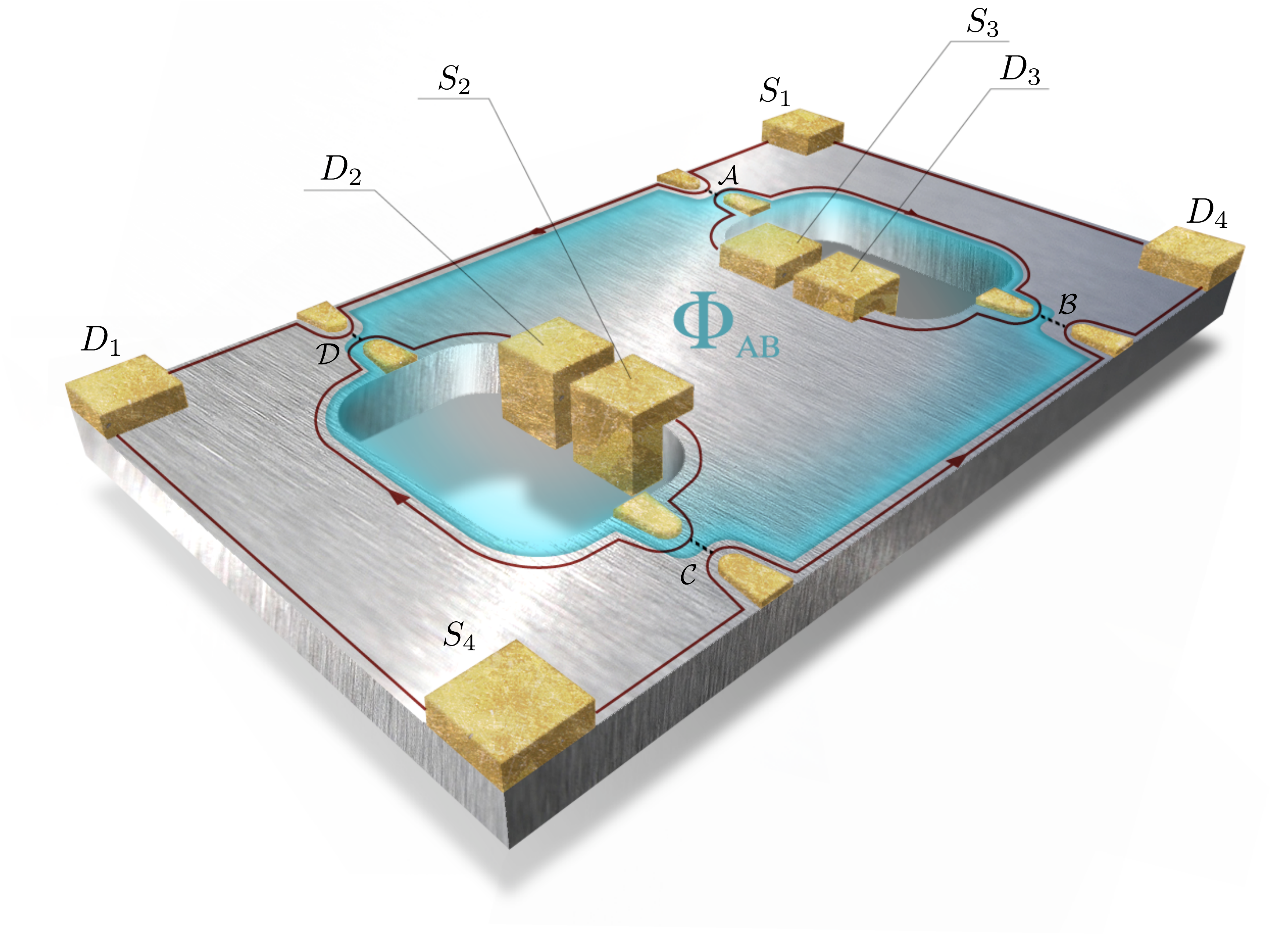}
\end{center}
\caption{
Scheme of a Hanbury Brown and Twiss setup realized with edge states. The relevant edges for our analysis are the lines $\overline{S_iD_i}$ with i=1,2,3,4. The external edges,
$\overline{S_1D_1}$ and  $\overline{S_4D_4}$,  are kept at potential $V$, the internal ones, $\overline{S_2D_2}$ and  $\overline{S_3D_3}$, are grounded. Inter-edge tunneling (dashed lines) takes place at the four QPCs, $\mathcal{A}$, $\mathcal{B}$, $\mathcal{C}$, $\mathcal{D}$, with tunneling amplitudes, $\Gamma_\mathcal{A}$, $\Gamma_\mathcal{B}$, $\Gamma_\mathcal{C}$, $\Gamma_\mathcal{D}$, respectively. The bias $V$ is assumed to be large compared to the
thermal energy, $k_B T$, so that only processes that involve transfer of QPs from the external to the internal edges are relevant for our analysis.
$L_i$ (not shown in the figure) is the distance between two consecutive QPCs along edge $\overline{S_iD_i}$. The magnetic flux threading the blue area, $\Phi_{\text{tot}}$, is relevant for the interferometry discussed here.
 } \label{HBT}
\end{figure}

Our main analysis, outlined below, reinforces the observation  that  the scattering  of two Laughlin anyons  is bosonic-like.  At the same time,  it also reveals  the non-analytic  structure of the interferometry signal of such anyons,  implying that the latter  are not  simple interpolation between fermions and bosons. The schematic setup is depicted in Fig.~\ref{HBT}. What replaces optical beams in the solid state device are edge states of the quantum Hall effect, formed due to the presence of strong perpendicular magnetic field. The chirality of these edge states allows the transport of charge excitations over large distances. The presence of an Aharonov-Bohm (AB) flux ($\Phi_{\text{AB}}$) provides an important handle to control and analyze the HBT correlations. The analogue of half-silvered mirrors are quantum point contacts (QPCs), which facilitate controlled forward transmission/backscattering reflection of the impinging charge excitation. Electron interferometers with such features have been realized~\cite{Yang:2003,Neder:2007}. We focus on the magnetic flux sensitive part of the current-current correlation, and show that the result is radically different from what has been predicted~\cite{Samuelsson:2004}, and later observed~\cite{Neder:2007} for the electronic case $(\nu=1)$.

A QP in a quantum Hall liquid at Laughlin filling factor $\nu$ can be described as a composite object, consisting of a point charge $q=\nu e$ with a single quantum magnetic flux solenoid, $\Phi_0=hc/e$, attached to it. When a QP encircles another QP it will pick up an AB-phase $\theta= 2 \pi \nu$ which accounts for their mutual fractional statistics~\cite{Arovas:1984}. When a QP tunnels from the external to the internal edges, its flux is {\em trapped} inside the interferometer~\cite{Kane:2003,Law:2006}. The magnetic flux enclosed in the active area of the interferometer (depicted in blue in Fig.~\ref{HBT}) is
$\Phi_{\text{tot}}(n)=\Phi_{AB}+\Phi_{stat}(n)$, where $\Phi_{stat}$ is the {\em statistical flux} and is given by $\Phi_0$ times the number, $n$, of trapped QPs. The dynamics of QPs moving along the edges of the interferometer is then entirely determined by $n$ mod$(3)$. i.e. for a given value of
 $\Phi_{AB}$ the system can be found in three possible {\em states} characterized by $n=0,1,2$.

For the study of the {\em non-equilibrium} dynamics of our strongly interacting HBT interferometer, we address the Markovian  evolution of the system among the three possible values of the statistical flux. Our microscopic Keldysh analysis simplifies, and can be cast in terms of rate equations for a certain parameter range~\cite{param_range}. The rate equations (whose coefficients are obtained by a microscopic analysis) carry information on interference effects of current cross correlations. Below we treat the QP tunneling current at each QPC perturbatively.

 Let us define the quantities needed in the ensuing analysis:
 $\ave{ I_i} $ is the average tunneling current  measured in drain $i$ and $S_{i,i^\prime}\equiv\int_{-\infty}^{\infty} dt \ave{ \left(I_i(t)-\ave{I_i}\right)\left(I_{i^\prime}(0)-\ave{I_{i^\prime}}\right)}$ is the zero-frequency current-current correlations between drains $i$ and $i^\prime$. The latter is the main object of this Letter. Next, we define $P(f,t|j)$, the probability to find the system with statistical flux $f$ [indices are defined mod$(3)$] at time $t$ given that it had statistical flux $j$ at time zero.
The system's dynamics is governed by a standard Master equation
\begin{align}
	\frac{d}{dt}P(f,t|j)=\sum_{k =0,1,2}\left[P(k,t|j)W_{k,f}-P(f,t|j)W_{f,k}\right]\, .
	\label{master}
\end{align}

Here $W_{j,f}$ is the total transition rate from the state $j$ to the state $f$.
In order to study the magnetic flux dependent part of the current-current correlations, we need to consistently include at least single-QP processes and two-QP processes, i.e. second and fourth order in the tunneling amplitudes $\Gamma$, respectively. In the limit of high voltage bias, $eV\gg k_B T$, considered here, only processes that transfer QPs from the outer to the inner edges are relevant.

Several microscopic processes, labeled by $\zeta$, contribute to each $W_{j,f}$ such that $W_{j,f}=\sum_\zeta W_{j,f}^{(\zeta)}$. The processes allowed are either $W_{j,j}^{(\zeta)}$, $W_{j,j+1}^{(\zeta)}$, or $W_{j,j+2}^{(\zeta)}$. The former renormalizes the vacuum current and does not affect any quantity calculated below. $W_{j,j+1}^{(\zeta)}$ has contributions from single QP processes (independent of flux, hence, independent of $j$), as well as from two-QPs processes (dependent of flux). $W_{j,j+2}^{(\zeta)}$  consists of two-QPs processes, and may or may not be flux-dependent. Each of the rates can be written as $W_{j,f}^{(\zeta)}=\tilde{W}^{(\zeta)}\kappa_j^{(\zeta)}$ with $\kappa_j^{(\zeta)}$ discussed in the caption of Table~\ref{table1}, which depicts all relevant processes.

\begin{table}[!ht]
\begin{tabular}{|l|c|c|r|c|c|r|}
 \hline
 \multicolumn{7}{|c|}{Elementary processes} \\
 \hline
  Process $\zeta$          & Order    & $(j,f)$ & $D_1$    &   $D_2$  &  $D_3$   & $D_4$  \\
  \hline
  $(1,\mathcal{A},0)$  & $\Gamma^2$ & $(j,j+1)$ & $-1$ & $0$ & $1$ &  $0$ \\
  $(1,\mathcal{B},0)$  & $\Gamma^2$ & $(j,j+1)$ &  $0$ & $0$ & $1$ & $-1$ \\
  $(1,\mathcal{C},0)$  & $\Gamma^2$ & $(j,j+1)$ &  $0$ & $1$ & $0$ & $-1$ \\
  $(1,\mathcal{D},0)$  & $\Gamma^2$ & $(j,j+1)$ & $-1$ & $1$ & $0$ &  $0$ \\
  $(2,\mathcal{A},0)$  & $\Gamma^4$ & $(j,j+2)$ & $-2$ & $0$ & $2$ &  $0$ \\
  $(2,\mathcal{B},0)$  & $\Gamma^4$ & $(j,j+2)$ &  $0$ & $0$ & $2$ & $-2$ \\
  $(2,\mathcal{C},0)$  & $\Gamma^4$ & $(j,j+2)$ &  $0$ & $2$ & $0$ & $-2$ \\
  $(2,\mathcal{D},0)$  & $\Gamma^4$ & $(j,j+2)$ & $-2$ & $2$ & $0$ &  $0$ \\
  $(2,\mathcal{A}$$\mathcal{B},0)$ & $\Gamma^4$ & $(j,j+2)$ & $-1$ & $0$ & $2$ & $-1$ \\
  $(2,\mathcal{C}$$\mathcal{D},0)$ & $\Gamma^4$ & $(j,j+2)$ & $-1$ & $2$ & $0$ & $-1$ \\
  $(2,\mathcal{A}$$\mathcal{D},0)$ & $\Gamma^4$ & $(j,j+2)$ & $-2$ & $1$ & $1$ &  $0$ \\
  $(2,\mathcal{B}$$\mathcal{C},0)$ & $\Gamma^4$ & $(j,j+2)$ &  $0$ & $1$ & $1$ & $-2$ \\
  $(2,\mathcal{A}\mathcal{B}\mathcal{C}\mathcal{D},\Phi_{\text{tot}}(j))$        & $\Gamma^4$ & $(j,j+2)$ & $-1$ & $1$ & $1$ & $-1$ \\
  $(1,\mathcal{A}\mathcal{B}\mathcal{C}\mathcal{D},\Phi_{\text{tot}}(j))_{1}$    & $\Gamma^4$ & $(j,j+1)$ &  $0$ & $0$ & $1$ & $-1$ \\
  $(1,\mathcal{A}\mathcal{B}\mathcal{C}\mathcal{D},\Phi_{\text{tot}}(j))_{2}$    & $\Gamma^4$ & $(j,j+1)$ & $-1$ & $1$ & $0$ &  $0$ \\
  \hline
\end{tabular}
\caption{Elementary QP transfer processes. Each process, $(\zeta)=(m,N,\phi)$, is characterized according to the change, $m$, in the number of QPs trapped in the interferometer; $N$ the QPCs at which QP tunneling takes place; and the flux, $\phi$, entering the flux factor $\kappa_j^{(m,N,\phi)}=\cos[2\pi \phi/(3\Phi_0)]$. Note that $\phi=0$ depicts a flux-independent process and $\phi=\Phi_{\text{tot}}(j)=\Phi_{\text{AB}}+j\cdot\Phi_0$ a process that depends on the total trapped flux. The order of the process (second or fourth in the tunneling amplitude $\Gamma$), the initial and final fluxon states [$(j,f)$, where $f-j$ is the added number of statistical fluxons], and the charge added at each drain ($+1$ refers to the absorption of one QP or charge $q=-(1/3)|e|$ at the drain), are indicated. For example (cf. Fig.~\ref{processes}), the process $(\zeta)=(1,\mathcal{A},0)$ corresponds to the emission of a QP from source $S_1$, its tunneling across QPC $\mathcal{A}$, and its trapping at $D_3$. Following the tunneling event a quasi-hole is created at edge $\overline{S_1 D_1}$ and a charge $-q$ is consequently absorbed in $D_1$. The flux dependent processes [the two-QPs trapping process $(2,\mathcal{A}\mathcal{B}\mathcal{C}\mathcal{D},\Phi_{\text{tot}}(j))$ and the single-QP trapping $(1,\mathcal{A}\mathcal{B}\mathcal{C}\mathcal{D},\Phi_{\text{tot}}(j))_{1}$] are illustrated in Fig.~\ref{processes}.
} \label{table1}
\end{table}

\begin{figure}
\includegraphics[width=8cm]{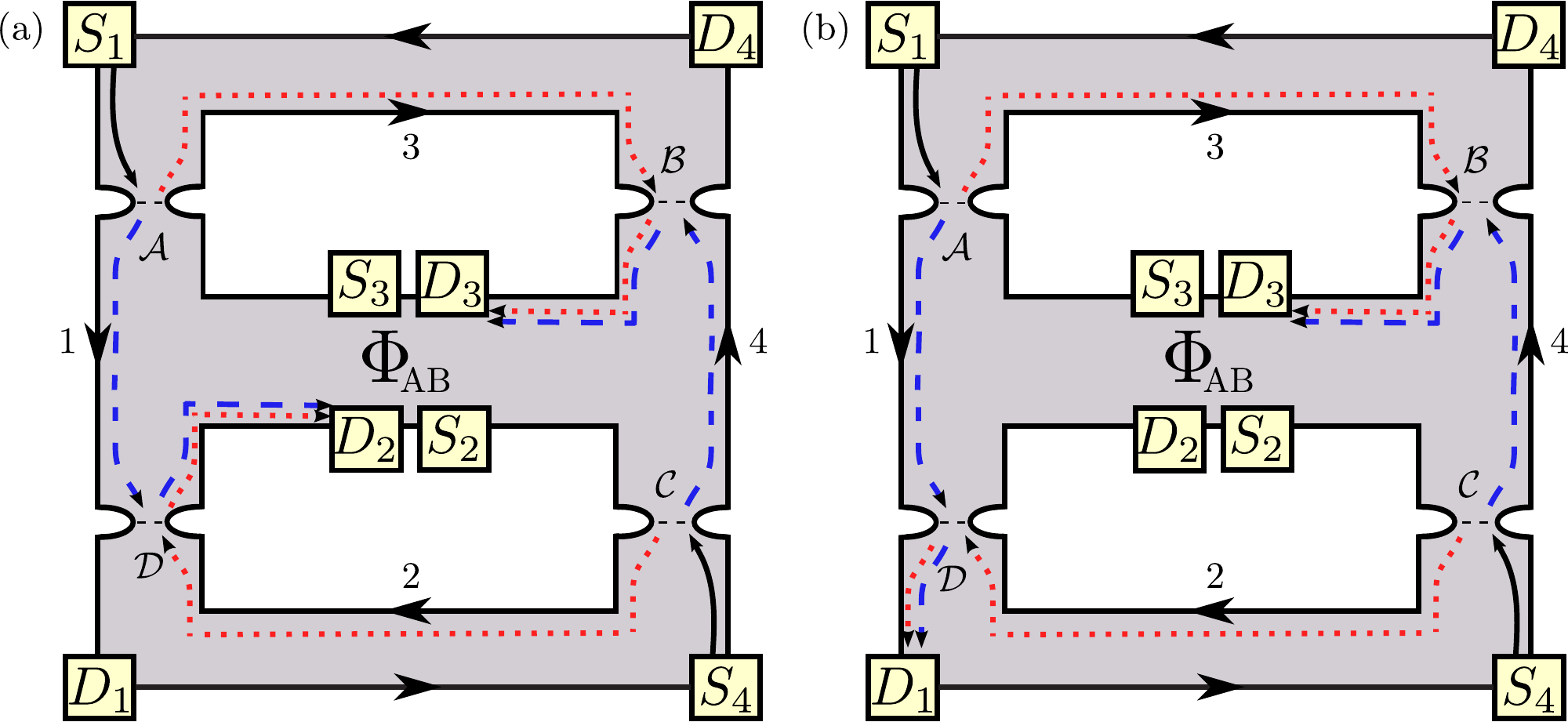}
\caption{(a) In process $(2,\mathcal{A}\mathcal{B}\mathcal{C}\mathcal{D},\Phi_{\text{tot}}(j))$ two QPs are transferred from edges 1 and 4 to edges 2 and 3, the process is AB-sensitive due to the interference between two amplitudes $A_1$ and $A_2$. In $A_1$ a QP tunnels from edge 1 to edge 3 and a second QP tunnels from edge 4 to edge 2 (red dotted line). In $A_2$ a QP tunnels from edge 1 to edge 2 and a second QP tunnels from edge 4 to edge 3 (blue dashed line). This process changes the statistical flux by two. (b) Process $(1,\mathcal{A}\mathcal{B}\mathcal{C}\mathcal{D},\Phi_{\text{tot}}(j))_1$ (and similarly process $(1,\mathcal{A}\mathcal{B}\mathcal{C}\mathcal{D},\Phi_{\text{tot}}(j))_2$) is also AB-sensitive but in this case only one QP is trapped inside the interferometer changing the statistical flux by one.}
\label{processes}
\end{figure}

Consider, first, the current collected at any of the drains. Assuming short-range interactions~\cite{wenBook}, which is reasonable in the presence of a metallic top gate, this current is flux independent (similarly to the $\nu=1$ case~\cite{Samuelsson:2004}), hence, is not of interest for us here. The following argument can be used to show this: consider for instance the current at drain $D_3$: owing to
the chiral propagation along the edges this tunneling current does not depend on the scattering at QPC $\mathcal{D}$. A gauge transformation can then ascribe the total magnetic flux to QPC $\mathcal{D}$ --- hence the current in drain 3 is AB independent. A similar argument holds for the tunneling currents collected at the other drains.


We, next, consider the AB-dependent component of the cross-current correlations. It is sufficient to express the following rates: the single-QP rates $W^{(1,\mathcal{A}),0}_{j,j+1}=\gamma |\Gamma_{\mathcal{A}}|^2$ (and similar expressions for rates involving the processes $(1,\mathcal{B},0)$, $(1,\mathcal{C},0)$, and $(1,\mathcal{D},0)$); and the two-QPs rate
\[
W_{j,j+2}^{(2,\mathcal{A}\mathcal{B}\mathcal{C}\mathcal{D},\Phi_{\text{tot}}(j))}=\Omega |\Gamma_{\mathcal{A}}\Gamma_{\mathcal{B}} \Gamma_{\mathcal{C}} \Gamma_{\mathcal{D}}| \cos[\frac{2\pi}{3}
(\Phi_{\text{AB}}+j\cdot\Phi_0)/\Phi_0 ].
\]
Here the $\Gamma$'s are the QPs tunneling amplitudes at the four QPCs, $\gamma$ and $\Omega$ are coefficients to be calculated
below. Using the method developed in references~\cite{Korotkov:1994,Koch:2006} we are able to calculate the AB-dependent component
of the cross-current correlator $S_{1,4}$:
\begin{align}\label{result}
	S^{\text{AB}}_{1,4}&\equiv S_{1,4}-\ave{S_{1,4}}_{\text{AB}}\\
	&= \frac{e^2|\Gamma_\mathcal{A}\Gamma_\mathcal{B}\Gamma_\mathcal{C}\Gamma_\mathcal{D}|^3\Omega^3 \cos[2\pi(\Phi_{\text{AB}}/\Phi_0)]}{6 (|\Gamma_\mathcal{A}|^2+|\Gamma_\mathcal{B}|^2+|\Gamma_\mathcal{C}|^2+|\Gamma_\mathcal{D}|^2)^2\gamma^2}\, ,\nonumber
\end{align}
where $\ave{\,}_{\text{AB}}$ refers to averaging over $\Phi_{\text{AB}}$.




{\em Model and Methods}---The low energy physics of the system is well described by the effective bosonic Hamiltonian~\cite{wenBook}
\begin{equation}
H_0=\frac{\hbar v}{4\pi}\sum_{l=1}^4\int dx (\partial_x\phi_l)^2\, ,
\end{equation}
describing chiral plasmonic excitations on the four edges (e.g. $\overline{S_1 D_1}$) of the interferometer. Here, $v$ is the plasmonic velocity at the edge. The bosonic fields $\phi_l$ satisfy the commutation relations are $\left[  \phi_l(x,t=0),\phi_k(x',t=0) \right]=i \pi \delta_{lk}\mbox{sgn}(x-x').$
The operators $\mbox{exp}(i \phi_l/\sqrt{\nu} )$ and $\mbox{exp}(i \sqrt{\nu} \phi_l )$ are, respectively, proportional to the electron
and the quasiparticle creation operator on the edge $l$.

To fully account for the quantum statistics of such particles one needs to multiply these bosonic operators by ``string operators'', known as Klein factors~\cite{vonDelft:1998,Kane:2003,Vishveshwara:2003,Law:2006}. In our analysis this procedure is replaced by carefully accounting for the dynamics of the statistical flux, attached to the tunneling QPs.

The total Hamiltonian, $H=H_0+H_T$, includes a tunneling part, $H_T=(H_T^{\mathcal{A}}+H_T^{\mathcal{B}}+H_T^{\mathcal{C}}+H_T^{\mathcal{D}})+h.c.$, which accounts for the most relevant tunneling operators at the QPCs. We assume that the external (internal) edges are tuned at voltage $V$ ($0$), select a gauge whereby the flux dependence is attached to
$H_T^{\mathcal{A}}$, and redefine the vacuum value of the fields at the external edges $\phi_l (x)\rightarrow \phi_l (x)-eV\sqrt{\nu} x/(v \hbar)$ ($l=1,4$). With these manipulations the tunneling operators read
\begin{align*}
H_T^{\mathcal{A}}(t,n)=&\Gamma_\mathcal{A} e^{i e \nu Vt/\hbar}e^{2\pi i \nu(\Phi_{ab}+\Phi_n)/\Phi_0 } e^{i\sqrt{\nu}(\phi_1(0,t)-\phi_3(0,t))} ,\\
\\
H_T^{\mathcal{B}}(t)=&\Gamma_\mathcal{B} e^{i e \nu Vt/\hbar}e^{-i e \nu V L_4/(\hbar v)}   e^{i\sqrt{\nu}(\phi_4(L_4,t)-\phi_3(L_3,t))}, \\
\\
H_T^{\mathcal{C}}(t)=&\Gamma_\mathcal{C}e^{i e \nu Vt/\hbar}   e^{i\sqrt{\nu}(\phi_4(0,t)-\phi_2(0,t))}, \\
\\
H_T^{\mathcal{D}}(t)=&\Gamma_\mathcal{D}e^{i e \nu Vt/\hbar} e^{-i e \nu V L_1/(\hbar v)}    e^{i\sqrt{\nu}(\phi_1(L_1,t)-\phi_2(L_2,t))} . \\
\end{align*}
Note that the magnetic flux attached to $H_T^{\mathcal{A}}$ comprises of both the AB-flux and the statistical flux due to $n$ $mod(3)$ QPs.



We next calculate the transition rates. The above model facilitates the calculation of the rates of the processes appearing in Table~\ref{table1}.
Rates are computed using generalized Fermi's golden rule (see, e.g.~\cite{Bruus:2004}) in order to evaluate single and two particles transfer between the edges. Generally we can write the transition rate between any initial state $|\psi_i\rangle$ with thermal  occupation $\rho_i$ to any final state $|\psi_f \rangle$ obtained from the initial one by transferring one or two QPs as $W_{i \rightarrow f}^{(\zeta)}=(2\pi/\hbar) \sum_{if}\rho_i |\ave{\psi_i| \tilde{T}|\psi_f}|^2\delta(E_f-E_i)$, where $\tilde{T}=H_T+H_T(E_i-H_0-i0^+)^{-1}H_T+\cdots $.
For example,
\begin{small}
\begin{multline}
W^{(2,\mathcal{A}\mathcal{B}\mathcal{C}\mathcal{D},\Phi_{\text{tot}}(j))}_{j,j+2}= |\Gamma_{\mathcal{A}}\Gamma_{\mathcal{B}} \Gamma_{\mathcal{C}} \Gamma_{\mathcal{D}}| \cos[\frac{2\pi}{3}
(\Phi_{\text{AB}}+\Phi_j)/\Phi_0 ] \times
\\
\int_{-\infty}^{\infty}\frac{d\epsilon}{2 \pi} G_{4}^{<}(\epsilon,-L_4)G_{3}^{>}(\epsilon+\nu e V,L_3)G_{2}^{>}(\epsilon+\nu e V,L_2)G_{1}^{<}(\epsilon,-L_1)
\end{multline}
\end{small}
Here we have introduced $G_i^{>}(\epsilon,x)$ and $G_i^{<}(\epsilon,x)$, the Green's functions in energy-space representation. In time-space representation they are given by $G_i^{>}(t,x)=\langle  e^{-i \sqrt{\nu}\phi_i(x,t)} e^{i \sqrt{\nu}\phi_i(0,0)} \rangle$ and $G_i^{<}(t,x)=\langle  e^{i \sqrt{\nu}\phi_i(0,0)} e^{-i \sqrt{\nu}\phi_i(x,t)} \rangle $.
We find for $\gamma$ and $\Omega$ (cf. Eq.~\eqref{result})
\begin{multline}
\gamma=\tilde{C}
\frac{ \beta^{1/3}e^{\pi \alpha/2}}{[1+2 \cosh(\pi \alpha)]\Gamma(\frac{2}{3})\Gamma(\frac{2}{3}-\frac{i\alpha}{2})
\Gamma(\frac{2}{3}+\frac{i\alpha}{2})}, \label{single-rate}
\end{multline}
and
\begin{equation}\label{2-phi-rate}
\Omega= \frac{32 \pi^{1/3}l_c^{4/3}\beta^{5/3}}{2^{1/3}\Gamma(2/3)\alpha^{4/3}\hbar^{7/3}v^{4/3}}
\,,
\end{equation}
where $\alpha=\nu e V/(\pi k_B T)$, $\beta=1/(k_B T)$, and $\tilde{C}=(l_{c}/v)^{2/3} (2 \pi/\hbar)^{5/3}$. In order to obtain Eq.(\ref{2-phi-rate}) we have assumed that the lengths of the individual interferometer arms satisfy $L_1+L_4-L_2-L_3 \ll \hbar v/(\nu e V)$ and $\alpha \gg1$.

\textit{Discussion ---} Eq.~\eqref{result} is our main result. We first note that the leading AB cross-current dependence comes with a plus sign, akin to bosonic HBT correlations (cf. refs. \onlinecite{Vishveshwara:2003,Samuelsson:2009}). This conclusion is in qualitative agreement with our toy model discussed above. The structure of Eq.~\eqref{result} is  worth noting as well. For an electronic two-particle interferometer operating in the integer quantum Hall regime ($\nu=1$), the leading flux-dependent contribution in the weak tunneling regime
is proportional to~\cite{Samuelsson:2004}  $\Gamma^4$ (bosons would behave the same way). Likewise one might expect the fundamental flux periodicity of $S^{AB}_{1,4}$ to be $\Phi_0$, in line with gauge invariance~\cite{Byers:1961,Kane:2003,Law:2006}. This would suggest that $S^{AB}_{14}$ is proportional to $\Gamma^{12}$, representing a coherent sequence of three $(2,\mathcal{A}\mathcal{B}\mathcal{C}\mathcal{D},\Phi_{\text{tot}}(j))$ two-QP processes. Our result for $S^{AB}_{14}$ scales as $\Gamma^{8}$, implying that an expansion of $S^{AB}_{14}$ in two-QPs rates, $W_{j,j+2}^{(2,\mathcal{A}\mathcal{B}\mathcal{C}\mathcal{D},\Phi_{\text{tot}}(j))}$, is non-analytic. This unique scaling with $\Gamma$ is the signature of QP HBT interference. Formally, this intriguing behavior is the outcome of the dressing of two-QPs processes by an infinite series of single-QP processes. We notice that the above results applies to the case of $\nu=1/3$ considered here; for a generic Laughlin filling factor $\nu=1/m$ ($m$ odd), having $m$ possible values of the statistical flux results in $S^{AB}_{1,4}$ being proportional to $\Gamma^{(2m+2)}$.

In summary, we have found that the scattering of two uncorrelated anyonic beams gives rise to HBT correlations which are bosonic in nature. This has been shown for a HBT interferometer threaded by an Aharonov-Bohm flux, and has also been suggested by the analysis of our toy model. The amplitude of the flux dependent cross-current correlations is non-analytic in the rates of the elementary two-anyon processes. Generalizing our model to finite temperatures ($eV\sim k_B T$) allows QPs to tunnel from the inner edges to the outer edges, but otherwise no quantitative changes are expected. The extension to finite range interaction will introduce higher harmonics at the flux dependence~\cite{Chalker:2007}. More interesting is the inclusion of multi-channel edges (going beyond Laughlin fractions), and eventually the generalizations to QPs satisfying non-Abelian statistics.

{\em Acknowledgments}---We thank A. Carmi, I. Protopopov and H.-S. Sim for useful discussions; and N. Gontmakher for the illustration of the device. We acknowledge financial support by BSF under Grant No. 2006371, GIF, ISF, Israel-Korea MOST grant, DFG CFN, and NSF under Grant No. DMR-0544116.


\section*{Supplementary Material}
Here we present a derivation of Eq.(\ref{single-rate}) and Eq.(\ref{2-phi-rate}), the one-particle and two-particle
rates, respectively. Let $| \psi_i \rangle$ and $|\psi_f \rangle$ be two many-body eigenstates of the system  in absence of tunneling (the tunneling Hamiltonian is $H_T$). Very generally the transition rate between them
due to the tunneling Hamiltonian  can be written as
\begin{equation}\label{generalfermigr}
\frac{2\pi}{\hbar}  |\langle\psi_i| \tilde{T}|\psi_f\rangle|^2 \delta(E_f-E_i) \, ,
\end{equation}
where $\tilde{T}$ is the scattering matrix given by
\begin{equation}
\tilde{T}=H_T+H_T\frac{1}{E_i-H_0-i 0^+}H_T+\cdots \, .
\end{equation}
Let us first consider the case of one-particle rate. For the sake of concreteness, we consider here tunneling through QPC
$\mathcal{A}$, all the other single particle rates being similar. In this case $|\psi_f \rangle$ is
obtained by removing a QP from edge 1 and transferring it to edge 3. Since we are interested in the {\em total} transition rate, we sum over all possible initial and final states. Notice that each  edge is kept at a finite chemical potential $\mu_i$ ($i=\{1,2,3,4\}$) and that the initial states  are weighted  by $w_i=Z^{-1}\langle \psi_i|\exp{[-\beta(H_0-\sum_i\mu_i N_i)}]|\psi_i\rangle$, with $Z=\mbox{Tr}\exp{[-\beta(H_0-\sum_i\mu_i N_i)]}$.
To the lowest order in the tunneling amplitude the transition rate $W^{(1,\mathcal{A},0)}_{j,j+1}$ is given by
\begin{equation}
\label{singletransfer}
W_{j,j+1}^{(1,\mathcal{A},0)}=\frac{2 \pi}{\hbar} \sum_{i,f} w_i \langle \psi_i | H_T^\mathcal{A} |\psi_f \rangle \langle \psi_f |H_T^{ \mathcal{A} \dag} | \psi_i \rangle \delta(E_f-E_i) \,.
\end{equation}
Here the operator $H_T^{ \mathcal{A} \dag}$ annihilates a quasiparticle on edge 1 and creates it on edge 3.  Expressing the tunneling operators in the interaction representation (with respect to $H_0$), Eq.(\ref{singletransfer}) can be  rewritten as
\begin{equation}\label{single-qp-transf2}
W_{j,j+1}^{(1,\mathcal{A},0)}= \sum_{if}   w_i \int_{-\infty}^{\infty}dt  \langle \psi_i | H_T^\mathcal{A}(0)|\psi_f \rangle \langle \psi_f | H_T^{ \mathcal{A} \dag}(t) | \psi_i \rangle \,.
\end{equation}
Notice that in Eq.(\ref{single-qp-transf2})  we can extend the sum over final states to a sum over a complete set of states and obtain
\begin{multline}\label{single-qp-transf3}
W_{j,j+1}^{(1,\mathcal{A},0)}=\int_{-\infty}^{\infty}dt
 \langle   H_T^\mathcal{A}(0) H_T^{ \mathcal{A} \dag}(t)  \rangle
\\ =|\Gamma_\mathcal{A}|^2l_{c}^{2/3}\int_{-\infty}^{\infty}dt\, e^{-i\nu e V t/\hbar}\left\{ \frac{\beta \hbar v}{\pi}
\sin \left[ \frac{\pi}{\beta \hbar v} (-i v t+l_c)   \right] \right\}^{-2/3}
\end{multline}
The integration leads to Eq.(\ref{single-rate}). In order to obtain Eq.(\ref{single-qp-transf3}) we note that the two point  correlation function for an edge kept at finite temperature and finite chemical potential is (this is equivalent to what presented in the main text)
\begin{small}
\begin{multline}\label{finite-mu-corr}
\langle e^{i \sqrt{\nu}\phi(x,t)} e^{-i \sqrt{\nu}\phi(0,0)} \rangle_\mu=   e^{i\mu \nu (t-x/v)/\hbar}
\langle e^{i \sqrt{\nu}\phi(x,t)} e^{-i \sqrt{\nu}\phi(0,0)} \rangle_{\mu=0}
\\
=e^{i\mu \nu (t-x/v)/\hbar} l_c^{1/3} \left\{\frac{\beta \hbar v }{\pi}\sin\left[ \frac{\pi}{\beta \hbar v}(-i(vt-x)+l_c)\right]\right\}^{-1/3}
\,.
\end{multline}
\end{small}
Let us now consider the {\em total} rate of transferring two quasiparticles from the external to the internal edges. Since there are no contributions to such a rate from second  and third order terms in the tunneling amplitudes $\Gamma$s, we need to consider the fourth order, we thus have
\begin{multline}\label{total-2qp-rate}
W_{j,j+2}=\frac{2\pi}{\hbar} \sum_{if}w_i  \bra{\psi_i} H_T \frac{1}{E_i-H_0-i 0^+} H_T \ket{\psi_f}
\\ \times
\bra{\psi_f} H_T \frac{1}{E_i-H_0+i 0^+} H_T \ket{\psi_i}\delta(E_f-E_i)\, .
\end{multline}
Notice that in this case the many body eigenstate $\ket{\psi_f}$ is obtained from $\ket{\psi_i}$ by transferring two quasiparticles. Being interested only  in the lowest contribution to the current-current correlation modulated by the magnetic flux, we study the contributions proportional to $|\Gamma_\mathcal{A}\Gamma_\mathcal{B}\Gamma_\mathcal{C}\Gamma_\mathcal{D}|$. We have
\begin{small}
\begin{multline}\label{2phi-2qp-rate}
W^{(2,\mathcal{A}\mathcal{B}\mathcal{C}\mathcal{D},\Phi_{\text{tot}}(j))}_{j,j+2}=\frac{2\pi}{\hbar} \sum_{if} w_i \left\{
 \langle \psi_i| H_T^\mathcal{B} \frac{1}{E_i-H_0-i 0^+} H_T^\mathcal{D}| \psi_f \rangle \right. \\+ \left.
  \langle \psi_i| H_T^\mathcal{D}  \frac{1}{E_i-H_0-i 0^+} H_T^\mathcal{B}| \psi_f \rangle
 \right\} \times \\
\left\{
 \langle \psi_f |H_T^\mathcal{A\dag} \frac{1}{E_i-H_0+i 0^+} H_T^\mathcal{C\dag} | \psi_i\rangle + 
 \right. \\ \left.
 \langle \psi_f |H_T^\mathcal{C\dag}  \frac{1}{E_i-H_0+i 0^+} H_T^\mathcal{A\dag} | \psi_i\rangle \right\} \\
 \times \delta(E_f-E_i)+c.c..
\end{multline}\end{small}
The above contribution corresponds to the rate $(2,\mathcal{A}\mathcal{B}\mathcal{C}\mathcal{D},\Phi_{\text{tot}}(j))$ of table \ref{table1}, the corresponding amplitudes are represented in Fig. \ref{processes}. Indeed the operator $\tad$ ($\tcd$)  annihilates a quasiparticle on edge 1 (on edge 4) and then creates it on edge 2 (3) respectively; similar statements apply to the operators  $\tbd$ and  $\tdd$.
Let us consider one of the four contributions proportional to $\Gamma_\mathcal{A}^* \Gamma_\mathcal{B} \Gamma_\mathcal{C}^*\Gamma_\mathcal{D}$ obtained from Eq.(\ref{2phi-2qp-rate}),
\begin{multline}
\mbox{I}=\frac{2\pi}{\hbar} \sum_{if} w_i \langle \psi_i | \tb  \frac{1}{E_i-H_0-i 0^+} \td|  \psi_f \rangle \\
\times
 \langle  \psi_f |\tad  \frac{1}{E_i-H_0+i 0^+} \tcd|  \psi_i\rangle \delta(E_f-E_i).
\end{multline}
Once again, moving to the interaction representation one can  rewrite the previous expression as
\begin{multline}
\mbox{I}=\int_{-\infty}^{+\infty}dt \sum_{if}w_i \langle  \psi_i| \tb(0)  \frac{1}{E_i-H_0-i 0^+} \td(0)|  \psi_f \rangle
\\ \times
\langle  \psi_f |\tad(t)  \frac{1}{E_i-H_0+i 0^+} \tcd(t) |  \psi_i\rangle
\end{multline}
The sum over the final states may be changed to a sum over a complete set of states; we can rewrite the expression as
\begin{multline}
\mbox{I}=\int_{-\infty}^{+\infty}dt \sum_{i}w_i
 \\ \times
 \langle \psi_i| \tb(0)  \frac{1}{E_i-H_0-i 0^+} \td(0) \tad(t)  \frac{1}{E_i-H_0+i 0^+} \tcd(t) |\psi_i\rangle
\end{multline}
This may be rewritten as
\begin{multline}\label{contribution-I}
\mbox{I}=\int_{-\infty}^{+\infty}dt \int_{-\infty}^{0}dt_1\int_{-\infty}^{0} dt_2
\\ \times
\langle \tb(t_1)\td(0)\tad(t)\tcd(t+t_2)\rangle
\end{multline}
The other three contributions proportional to $\Gamma_\mathcal{A}^* \Gamma_\mathcal{B} \Gamma_\mathcal{C}^*\Gamma_\mathcal{D}$
clearly read:
\begin{multline}
\mbox{II}=\int_{-\infty}^{+\infty}dt \int_{-\infty}^{0}dt_1\int_{-\infty}^{0} dt_2
\\ \times
\langle \tb(t_1)\td(0)\tcd(t)\tad(t+t_2)\rangle \, ,
\end{multline}
\begin{multline}
\mbox{III}=\int_{-\infty}^{+\infty}dt \int_{-\infty}^{0}dt_1\int_{-\infty}^{0} dt_2
\\ \times
\langle \td(t_1)\tb(0)\tad(t)\tcd(t+t_2)\rangle \, ,
\end{multline}
\begin{multline}
\mbox{IV}=\int_{-\infty}^{+\infty}dt \int_{-\infty}^{0}dt_1\int_{-\infty}^{0} dt_2
\\ \times
\langle \td(t_1)\tb(0)\tcd(t)\tad(t+t_2)\rangle \, .
\end{multline}
We thus obtain
\begin{equation}
W_{j,j+2}^{(2,\mathcal{A}\mathcal{B}\mathcal{C}\mathcal{D},\Phi_{\text{tot}}(j))}=(\mbox{I}+\mbox{II}+\mbox{III}+\mbox{IV})+c.c.
\end{equation}
Again using Eq.(\ref{finite-mu-corr}) we can write Eq.(\ref{contribution-I}) as
\begin{multline}\label{rate}
\mbox{I}=\Gamma^*_\mathcal{A} \Gamma_\mathcal{B} \Gamma^*_\mathcal{C} \Gamma_\mathcal{D} e^{-2\pi i \nu(\Phi_{ab}+j\,\Phi)/\Phi_0 } e^{-i e \nu V (L_4+L_1)/v \hbar} \\ \times
\left(\frac{\pi l_c}{ \beta \hbar v}\right)^{4\nu}
\int_{-\infty}^{+\infty}dt\int_{-\infty}^{0}dt_1\int_{-\infty}^{0} dt_2 e^{-2i \nu e V t/\hbar} \\ \times
\sin \left\{ \frac{\pi}{\beta \hbar v}[i v (-t+\frac{t_1}{2}-\frac{t_2}{2})-i L_4+ l_c ]\right\}^{-\nu}
 \\ \times
 \sin \left\{ \frac{\pi}{\beta \hbar v}[i v (-t+\frac{t_1}{2}+\frac{t_2}{2})-i L_3+ l_c ]\right\}^{-\nu}
  \\ \times
\sin \left\{ \frac{\pi}{\beta \hbar v}[i v (-t-\frac{t_1}{2}-\frac{t_2}{2})-i L_2+ l_c ]\right\}^{-\nu}
  \\ \times
  \sin \left\{ \frac{\pi}{\beta \hbar v}[i v (-t-\frac{t_1}{2}+\frac{t_2}{2})-i L_1+ l_c ]\right\}^{-\nu}
\end{multline}
where $t$ has been shifted by $(t_1-t_2)/2$. Remarkably, changing variables in the terms II, III and IV, yields exactly the missing  sectors in the  $t_1$ and $t_2$ integrals of contribution I.
We can then combine the four contributions into a single expression
\begin{multline}\label{rate2}
\mbox{I+II+III+IV}=\Gamma^*_\mathcal{A} \Gamma_\mathcal{B} \Gamma^*_\mathcal{C} \Gamma_\mathcal{D} e^{-2\pi i \nu(\Phi_{ab}+j \,\Phi)/\Phi_0 }e^{-i e \nu V (L_4+L_1)/v \hbar} \\ \times
\left(\frac{\pi l_c}{ \beta \hbar v}\right)^{4\nu}
\int_{-\infty}^{+\infty}dt\int_{-\infty}^{0}dt_1\int_{-\infty}^{0} dt_2 e^{-2i \nu e V t/\hbar} \\ \times
\sin \left\{ \frac{\pi}{\beta \hbar v}[i v (-t+\frac{t_1}{2}-\frac{t_2}{2})-i L_4+ l_c ]\right\}^{-\nu}
 \\ \times
 \sin \left\{ \frac{\pi}{\beta \hbar v}[i v (-t+\frac{t_1}{2}+\frac{t_2}{2})-i L_3+ l_c ]\right\}^{-\nu}
  \\ \times
\sin \left\{ \frac{\pi}{\beta \hbar v}[i v (-t-\frac{t_1}{2}-\frac{t_2}{2})-i L_2+ l_c ]\right\}^{-\nu}
  \\ \times
  \sin \left\{ \frac{\pi}{\beta \hbar v}[i v (-t-\frac{t_1}{2}+\frac{t_2}{2})-i L_1+ l_c ]\right\}^{-\nu}
\end{multline}
This integral can be evaluated explicitly in the limit $\nu e V \beta >>1$ and $(L_1+L_4-L_2-L_3)<<\hbar v/(\nu e V)$ to obtain
\begin{multline}\label{2-phi-rate-bis}
W_{j,j+2}^{(2,\mathcal{A}\mathcal{B}\mathcal{C}\mathcal{D},\Phi_{\text{tot}}(j))}= 2 \left| \Gamma_\mathcal{A} \Gamma_\mathcal{B} \Gamma_\mathcal{C} \Gamma_\mathcal{D} \right| \cos \left[ 2\pi\nu(\Phi_\Gamma+\Phi_{ab}+j \,\Phi_0)/\Phi_0 \right] \\ \times  \frac{16 \pi^{1/3}l_c^{4/3}\beta^{5/3}}{2^{1/3}\Gamma(2/3)\alpha^{4/3}\hbar^{7/3}v^{4/3}}
\\ =  \Omega\left| \Gamma_\mathcal{A} \Gamma_\mathcal{B} \Gamma_\mathcal{C} \Gamma_\mathcal{D}  \right|  \cos \left[ 2\pi\nu(\Phi_\Gamma+\Phi_{ab}+j\,\Phi_0)/\Phi_0 \right]
   \, ,
\end{multline}
where we write:
\[
\Gamma^*_\mathcal{A} \Gamma_\mathcal{B} \Gamma^*_\mathcal{C} \Gamma_\mathcal{D} = \left| \Gamma_\mathcal{A} \Gamma_\mathcal{B} \Gamma_\mathcal{C} \Gamma_\mathcal{D} \right|
\exp(-i 2\pi \nu \Phi_\Gamma/\Phi_0) \, .
\]
Note that in the main text we take $\Phi_\Gamma$ to be zero for simplicity.
\end{document}